\RequirePackage{fix-cm}
\documentclass{svjour3}                     
\smartqed  
\usepackage{graphicx}
\usepackage{amssymb}
\usepackage{hyperref}
\usepackage{amsmath}
\usepackage{float}
\usepackage{hyperref}      
\journalname{Eur. Phys. J. Plus}
\begin{document}

\title{Constraining Bianchi type V universe with recent H(z) and BAO observations in Brans - Dicke theory of gravitation}


\titlerunning{Constraining Bianchi type V universe..}        

\author{R. Prasad$^{1}$, Avinash Kr. Yadav$^{2}$, Anil Kumar Yadav$^{3}$
}

\authorrunning{R. Prasad, Avinash Kr. Yadav and Anil Kumar Yadav} 

\institute{$^{1}$Department of Physics, Galgotias College of Engineering and Technology, Greater Noida - 201310, India, \email{drrpnishad@gmail.com}\\
$^{2}$Department of Mathematics, United College of Engineering and Research, Greater Noida - 201310, India \email{avinashyad75@gmail.com}\\
$^3$Department of Physics, United College of Engineering and Research, Greater Noida - 201310, India \email{abanilyadav@yahoo.co.in}}

\date{Received: 28 January 2020 / Accepted: 2 March 2020}

\maketitle
\begin{abstract}
In this paper, we investigate a transitioning model of Bianchi type V universe in Brans-Dicke theory of gravitation. The derived model not only validates Mach's principle but also describes the present acceleration of the universe. In this paper, our aim is to constrain an exact Bianchi type V universe in Brans - Dicke gravity. For this sake, firstly we obtain an exact solution of field equations in modified gravity and secondly constrain the model parameters by bounding the model with recent $H(z)$ and Baryon acoustic oscillations (BAO) observational data. The current phase of accelerated expansion of the universe is also described by the contribution coming from cosmological constant screened scalar field with deceleration parameter showing a transition redshift of about $z_{t} = 0.79$. Some physical properties of the universe are also discussed. \\
\end{abstract}

\textbf{Kewwords:} Bianchi V spaceptime; Brans-Dike gravity; Scalar field; Accelerating universe.\\

\textbf{Pacs:} 98.80.-k, 04.20.Jb, 04.50.kd\\

\section{Introduction}
\label{intro}
The supernovae Ia observations \cite{Riess/1998,Perlmutter/1999} have exhibited a strong evidence that our universe is dominated by two types of dark components at present epoch. These two components of present universe are named as dark matter and dark energy. Today, it is one of the major issues in modern cosmology to describe the nature of dark matter and dark energy. The dark matter has not been directly observed but there are many evidences such as galaxy rotation curves, gravitational effects, gravitational lensing etc which support the existence of dark matter. The dark energy is an unknown form of energy that pervades the whole universe. It is believed to have negative pressure, the dark energy is causing acceleration in the present universe. According to WMAP observations \cite{Bennett/2003,Hinshaw/2003,Spergel/2003}, the universe energy density appears to consist of approximately 4 \% of that of visible matter, 21 \% of that of dark matter and 75 \% of that of dark energy. In the literature, the acceleration in present universe is described by two ways i) inclusion of dark energy in right side of Einstein's equation i. e. by modifying energy-momentum tensor ii) modification in left side of Einstein's equation i. e. geometric modification. The authors of Refs. \cite{Copeland/2006,Bamba/2012} have described late time acceleration of the universe by considering dark energy and modified gravity respectively. Later on, numerous cosmological models have been investigated in General Relativity (GR) with inclusion of dark energy \cite{Akarsu/2010,Kumar/2011,Yadav/2011,Yadav/2011a,Yadav/2011b,Yadav/2016,Amirhashchi/2018,Amirhashchi/2017} and in modified theories of gravity without inclusion of dark energy \cite{Moraes/2017,Yadav/2014,Yadav/2018,Singh/2015,Myrzakulov/2012,Houndjo/2012,Kiani/2014,Yadav/2019}. Even after all these attempts, the reliable nature of dark energy has not been convincingly explained yet.\\ 

The Brans-dicke (BD) theory \cite{Brans/1961}, which is a natural generalization of GR, provides a worthy framework for dynamical dark energy models. In this this theory, the scalar field $\phi$ is being time-dependent and it is equivalent to $(8\pi G)^{-1}$. Therefore, in BD scalar-tensor theory, the scalar field $\phi$ couples to the gravity with a dimensionless coupling parameter $\omega$. It is worth to note that BD theory of gravitation commits expanding solutions for scalar field and average scale factor which are compatible with the solar system observations. In Refs. \cite{Bertolami/2000,Kim/2005,Clifton/2006}, the authors have investigated that BD theory explains the late time accelerated expansion of the universe and also conciliates the observation data. It is also to be noticed that BD theory of gravitation reduces to GR if scalar field is constant and $\omega \rightarrow \infty$ \cite{Rama/1996a,Rama/1996b}. Some new agegraphic dark energy models in Brans-Dicke gravity have been investigated \cite{Sheykhi/2010,Sheykhi/2011,Pasqua/2013,Fayaz/2016}. These models explain the late time accelerated expansion of the universe with evolution of scalar field as power law of scale factor. In the literature, BD theory is invoked to fulfill the requirement of Mach's principle\cite{Brans/1961,Fujii/2003,Faraoni/2004,Uehara/1982,Lorenz/1984}. In Sen and Sen \cite{Sen/2001}, authors have investigated that a perfect fluid cannot support acceleration but a fluid with dissipative pressure can drive late time acceleration of current universe. The present cosmic acceleration without resorting to a cosmological constant or quintessence matter has been investigated in BD theory but then Brans-Dicke coupling constant asymptotically acquires a small negative value for an accelerating universe at late time\cite{Banerjee/2001} while in Ref. \cite{Bertolami/2000}, authors have obtained solution for accelerating universe with $\phi^{2}$ potential for large BD coupling constant without considering positive energy condition for matter and scalar field both. Recently Akarsu et al. \cite{Akarsu/2020,Akarsu/2019} have investigated some particular negative range of $\omega$ and positive large value of $\omega$ that lead acceleration in massive Brans-Dicke gravity. Some large angle anomalies viewed in cosmic microwave background (CMB) radiations \cite{Spergel/2003} are favoring the presence of anisotropies in the early stage of the universe which violate the isotropical nature of the observable universe and hence to clearly describe the early universe - a spatially homogeneous but anisotropic Bianchi models play a significant role. In the literature, several Bianchi type models have been investigated with different matter distribution in Brans-Dicke theory of gravitation. In particular, Kiran et al \cite{Kiran/2015} have investigated an interacting Bianchi V cosmological model within the framework of Brans-Dicke cosmology. In the recent past, some Brans-Dicke anisotropic models have been studied to discuss the late time accelerated expansion of the universe \cite{Adhav/2014,Ramesh/2016,Reddy/2016,Naidu/2018}. Some useful applications of Bianchi type models compatible with astrophysical observations are given in Refs. \cite{Amirhashchi/2017,Akarsu/2019prd,Amirhashchi/2019a,Amirhashchi/2018,Goswami/2019mpla,Kumar/2011mpla}.\\

In this paper, we have investigated a Bianchi type V model of the universe filled with pressure-less matter and cosmological constant at present in Brans-Dicke gravity. Firstly we have obtained an exact Brans-Dicke universe and then find constraints on model parameters by using recent H(z) and BAO observational data. The rest of the paper is organized as follows: in section 2, we present the model and its basic equations. In section 3, we describe the method and likelihoods. In section 4, we discuss the physical and kinematic properties of the model under consideration. The summary of our findings is presented in section 5.       
            
\section{The model and Basic equations}\label{2}
The Einstein's field equations in Brans-Dicke theory is given by
\[
R_{ij}-\frac{1}{2}Rg_{ij}+\Lambda g_{ij}=\frac{8\pi}{\phi c^{2}}T_{ij}
\]
\begin{equation}
\label{BD-1}
-\frac{\omega}{\phi^{2}}\left(\phi_{i}\phi_{j}-\frac{1}{2}g_{ij}\phi_{k}\phi^{k}\right)-\frac{1}{\phi}(\phi_{ij}-g_{ij}\square\phi)
\end{equation}
and
\begin{equation}
\label{BD-2}
(2\omega+3)\square\phi=\frac{8\pi T}{c^{4}}+2\Lambda \phi
\end{equation}
where $\omega$ is the Brans-Dicke coupling constant; $\phi$ is Brans-Dicke scalar field and $\Lambda$ is the cosmological constant.\\ 

The Bianchi type V space-time is read as
\begin{equation}
\label{BI}
ds^{2}= dt^{2}- A(t)^{2}dx^{2}- e^{2 \alpha x} \left[B(t)^{2}dy^{2}+C(t)^{2}dz^{2}\right]
\end{equation}
where $A(t),~B(t)~\& C(t)$ are scale factors along $x$, $y$ and $z$ direction respectively and average scale factor is defined as $a = (ABC)^{\frac{1}{3}}$. The exponent $\alpha\neq0$ in \eqref{BI} is an arbitrary constant.\\
The energy momentum tensor of perfect fluid is given by
\begin{equation}
\label{em}
T_{ij}=(p+\rho)u_{i}u_{j}-pg_{ij}
\end{equation}
Here, $p$ and $\rho$ are the isotropic pressure and energy density of the matter under consideration. also $u^{i}u^{j} =-1$ and $u^{i}$ is the four velocity vector.\\
The field equations (\ref{BD-1}) for space-time (\ref{BI}) are read as
\begin{equation}
\label{ef-1}
\frac{\ddot{B}}{B}+\frac{\ddot{C}}{C}+\frac{\dot{B}\dot{C}}{BC}-\frac{\alpha^{2}}{A^{2}}+\frac{\omega}{2}\frac{\dot{\phi}^{2}}{\phi^{2}}+\frac{\dot{\phi}}{\phi}\left(\frac{\dot{B}}{B}+\frac{\dot{C}}{C}\right)+\frac{\ddot{\phi}}{\phi}=-\frac{8\pi p}{\phi}+\Lambda 
\end{equation} 
\begin{equation}
\label{ef-2}
\frac{\ddot{A}}{A}+\frac{\ddot{C}}{C}+\frac{\dot{A}\dot{C}}{AC}-\frac{\alpha^{2}}{A^{2}}+\frac{\omega}{2}\frac{\dot{\phi}^{2}}{\phi^{2}}+\frac{\dot{\phi}}{\phi}\left(\frac{\dot{A}}{A}+\frac{\dot{C}}{C}\right)+\frac{\ddot{\phi}}{\phi}=-\frac{8\pi p}{\phi}+\Lambda 
\end{equation}
\begin{equation}
\label{ef-3}
\frac{\ddot{A}}{A}+\frac{\ddot{B}}{B}+\frac{\dot{A}\dot{B}}{AB}-\frac{\alpha^{2}}{A^{2}}+\frac{\omega}{2}\frac{\dot{\phi}^{2}}{\phi^{2}}+\frac{\dot{\phi}}{\phi}\left(\frac{\dot{A}}{A}+\frac{\dot{B}}{B}\right)+\frac{\ddot{\phi}}{\phi}= -\frac{8\pi p}{\phi}+\Lambda 
\end{equation}
\begin{equation}
\label{ef-4}
\frac{\dot{A}\dot{B}}{AB}+\frac{\dot{B}\dot{C}}{BC}+\frac{\dot{C}\dot{A}}{CA}-\frac{3\alpha^{2}}{A^{2}}-\frac{\omega}{2}\frac{\dot{\phi}^{2}}{\phi^{2}}+\frac{\dot{\phi}}{\phi}\left(\frac{\dot{A}}{A}+\frac{\dot{B}}{B}+\frac{\dot{C}}{C}\right) = \frac{8\pi \rho}{\phi}+\Lambda
\end{equation}
\begin{equation}
\label{ef-4a}
2\dfrac{\dot{A}}{A}-\dfrac{\dot{B}}{B}-\dfrac{\dot{C}}{C} =0 \Rightarrow A^2 = BC
\end{equation}
\begin{equation}
\label{ef-5}
\frac{\ddot{\phi}}{\phi}+\left(\frac{\dot{A}}{A}+\frac{\dot{B}}{B}+\frac{\dot{C}}{C}\right)\frac{\dot{\phi}}{\phi} = \frac{8\pi(\rho-3p)}{(2\omega+3)\phi}+\frac{2\Lambda}{2\omega +3}
\end{equation}
where over dot denotes derivatives with respect to time t.\\

The equation of continuity is read as
\begin{equation}
\label{ef-6}
\dot{\rho}+(1+\gamma)\left(\frac{\dot{A}}{A}+\frac{\dot{B}}{B}+\frac{\dot{C}}{C}\right)\rho = 0
\end{equation}
where $\gamma$ is the equation of state parameter of perfect baro-tropic fluid and it is defined as $\gamma = \frac{p}{\rho} = constant$. The pressure of dark matter is zero which can be recover from baro-tropic equation of state by choosing $\gamma = 0$.  
\subsection{Solution of Einstein's field equations}
Equations (\ref{ef-1})-(\ref{ef-3}) lead the following system of equations 
\begin{equation}
\label{ef-7}
\frac{\ddot{A}}{A}-\frac{\ddot{B}}{B}+\frac{\dot{A}\dot{C}}{AC}-\frac{\dot{B}\dot{C}}{BC}+\left(\frac{\dot{A}}{A}-\frac{\dot{B}}{B}\right)\frac{\dot{\phi}}{\phi} = 0
\end{equation}
\begin{equation}
\label{ef-8}
\frac{\ddot{B}}{B}-\frac{\ddot{C}}{C}+\frac{\dot{A}\dot{B}}{AB}-\frac{\dot{A}\dot{C}}{AC}+\left(\frac{\dot{B}}{B}-\frac{\dot{C}}{C}\right)\frac{\dot{\phi}}{\phi} = 0
\end{equation}
\begin{equation}
\label{ef-9}
\frac{\ddot{C}}{C}-\frac{\ddot{A}}{A}+\frac{\dot{B}\dot{C}}{BC}-\frac{\dot{A}\dot{B}}{AB}+\left(\frac{\dot{C}}{C}-\frac{\dot{A}}{A}\right)\frac{\dot{\phi}}{\phi} = 0
\end{equation}
The equations (\ref{ef-7})-(\ref{ef-9}) are the system of three equations with four unknown variables $A$, $B$, $C$ and $\phi$. So, one can not solve these equations in general. In connection with equation (\ref{ef-4a}), one may propose the following relation among the metric functions
\begin{equation}
\label{s-1}
B = AD \;\;\; \& \;\; C = \frac{A}{D}
\end{equation}
where $D = D(t)$ measures the anisotropy in universe. For $D = 1$ and $\alpha = 0$, Bianchi V universe recovers the case of FRW universe.\\
Equations (\ref{ef-8}) and (\ref{s-1}) lead to
\begin{equation}
\label{s-2}
\frac{\ddot{D}}{D}-\frac{\dot{D}^{2}}{D^{2}}+\frac{\dot{D}}{D}\left(3\frac{\dot{A}}{A}+\frac{\dot{\phi}}{\phi}\right) = 0
\end{equation}
After integration of equation (\ref{s-2}), we obtain
\begin{equation}
\label{s-3}
D = exp\left[\int\frac{k}{A^{3}\phi}dt\right]
\end{equation} 
Now, the average scale factor is computed as 
\begin{equation}
\label{s-2}
a^{3} = ABC = A^{3} \Rightarrow a = A
\end{equation}

Therefore, the Friedmann equations (\ref{ef-1}) and (\ref{ef-4}) respectively recast as following
\begin{equation}
\label{ef-H1}
2\frac{\ddot{a}}{a}+\frac{\dot{a}^{2}}{a^{2}}-\frac{\alpha^{2}}{a^{2}}+\frac{k^{2}}{a^{6}\phi^{2}}+\frac{\omega}{2}\frac{\dot{\phi}^{2}}{\phi^{2}}+2\frac{\dot{\phi}}{\phi}\frac{\dot{a}}{a} + \frac{\ddot{\phi}}{\phi} = -\frac{8\pi p}{\phi}+\Lambda
\end{equation}
\begin{equation}
\label{ef-H2}
3\frac{\dot{a}^{2}}{a^{2}}-\frac{3\alpha^{2}}{a^{2}}-\frac{k^{2}}{a^{6}\phi^{2}}-\frac{\omega}{2}\frac{\dot{\phi}^{2}}{\phi^{2}}+3\frac{\dot{\phi}}{\phi}\frac{\dot{a}}{a} = \frac{8\pi \rho}{\phi}+\Lambda 
\end{equation}
\subsection{The model: Brans-Dicke anisotropic universe}
The density parameters are read as
\begin{equation}
\label{dp-1}
\Omega_{m} = \frac{8\pi \rho_{m}}{3 H^2 \phi},~~\Omega_{\Lambda}=\frac{\Lambda}{3H^2}, ~~ \Omega_{\sigma} = \frac{k^{2}}{3H^{2}a^{6}\phi^{2}},~~\Omega_{\alpha} = \frac{\alpha^{2}}{a^{2}H^{2}} 
\end{equation}
where $\rho_{m} = \left(\rho_{m}\right)_{0}a^{-3}$ is the energy density of pressure-less matter and $\Omega_{m}$, $\Omega_{\Lambda}$ and $\Omega_{\sigma}$ represent the dimensionless density parameters for dark matter, $\Lambda$- energy, shear anisotropy and $\alpha$ parameter respectively. $H$ is Hubble's parameter and it is defined as $H = \frac{\dot{a}}{a}$.\\

The deceleration parameter $q$ and scalar field deceleration parameter $q_{\phi}$ are read as\\
\begin{equation}
\label{dp-1a}
q = -\frac{\ddot{a}}{aH^{2}}, ~~~ q_{\phi} = -\frac{\ddot{\phi}}{\phi H^{2}}
\end{equation}
Dividing equations (\ref{ef-H2}) by $3H^{2}$ and then using equation (\ref{dp-1}), we have\\

\begin{equation}
\label{n-1}
\Omega_{m}+\Omega_{\Lambda} + \Omega_{\sigma} + \Omega_{\alpha}= 1+\psi-\frac{\omega}{6}\psi^{2}
\end{equation}
where $\psi = \frac{\dot{\phi}}{\phi H}$.\\
After some algebra in equations (\ref{ef-5}), (\ref{ef-H1}) and (\ref{ef-H2}), finally we obtained
\begin{equation}
\label{n-2}
\phi = \phi_{0}\left(\frac{a}{a_{0}}\right)^{\psi}\;\; \&\;\; \psi = \frac{1}{\omega+1}
\end{equation}
where $a_{0}$ is the present value of scale factor.\\
Thus, equation (\ref{n-1}) reduces to
\begin{equation}
\label{BV-1}
\Omega_{m}+\Omega_{\Lambda} + \Omega_{\sigma}+\Omega_{\alpha} = 1+\frac{5\omega+6}{6(\omega+1)^{2}}
\end{equation}
If we define the density of scalar field $\phi$ as
\begin{equation}
\label{Sf-1}
\Omega_{\phi}=-\frac{5\omega+6}{6(\omega+1)^{2}},
\end{equation}
then, equation (\ref{BV-1}) is recast as
\begin{equation}
\label{BV-2}
\Omega_{m}+\Omega_{\Lambda} + \Omega_{\sigma}+\Omega_{\alpha}+\Omega_{\phi} = 1
\end{equation}
The scale factor $a$ and $\phi$ in connection with $z$ are read as
\begin{equation}
\label{a-z}
a = \frac{a_{0}}{1+z},~~ \phi = \frac{1}{(1+z)^{\frac{1}{1+\omega}}}, 
\end{equation}  
Equations (\ref{dp-1}), (\ref{BV-2}) and (\ref{a-z}) leads to 
\begin{equation}
\label{H-BD}
H_{\sigma BD}= \frac{H_{0}}{(1-\Omega_{\phi})^\frac{1}{2}}\left[\Omega_{m0}(1+z)^{2.5 \left(\sqrt{1-\Omega \phi }-1\right)+3}+\Omega_{\sigma 0}(1+z)^{5 \left(\sqrt{1-\Omega \phi }-1\right)+6}+ \Omega_{\alpha}(1+z)^{2}+\Omega_{\Lambda 0}\right]^\frac{1}{2}
\end{equation}
where $H_{0}$, $\Omega_{m0}$, $\Omega_{\sigma 0}$ and $\Omega_{\Lambda 0}$ denote present values of Hubble constant and densities parameters due to dark matter, anisotropy and cosmological constant respectively.\\   
\section{Method and Likelihoods}\label{3}
In this section, we briefly describe the observational data and the statistical methodology to constrain the Bianchi V universe as discussed in the previous section.
\begin{itemize}
\item {\bf Observational Hubble Data (OHD)}: We adopt $46~H(z)$ datapoints over the redshift range of $0\leq z\leq 2.36$ obtained from cosmic chronometric (CC) technique. We have compiled all $46 H(z)$ datapoints in table \ref{tab:1}.\\
\item {\bf Baryon acoustic oscillations (BAO)}: We use 10 baryon acoustic oscillations data extracted from the 6dFGS \cite{Beutler/2012}, SDSS-MGS \cite{Ross/2015}, BOSS \cite{Alam/2017}, BOSS CMASS \cite{Anderson/2014}, and WiggleZ \cite{Kazin/2014} surveys.
\end{itemize}
\begin{table}[ht]
\caption{Hubble parameter versus redshift data.}
\centering
\setlength{\tabcolsep}{18pt}
\scalebox{0.98}{
\begin{tabular} {ccccc}
\hline
\hline
S.N. &z &H(z)$[Gyr^{-1}]$ &$\sigma_{i}$ $[Gyr^{-1}]$& References\\[0.5ex] 

\hline{\smallskip}
1 & 0  & 0.069 & 0.0013  & \cite{Macaulay/2018} \\
2 & 0.07 & 0.069 &  0.020 & \cite{Zhang/2014}\\
3 & 0.09 & 0.071 &  0.012 &  \cite{Simon/2005} \\
4 &0.01 & 0.071 &  0.012 &  \cite{Stern/2010} \\
5 &0.12 & 0.071 &  0.027 &  \cite{Zhang/2014}\\
6 &0.17 & 0.07 &  0.0081 &  \cite{Stern/2010} \\
7 &0.179 &0.085 &  0.0041 &  \cite{Moresco/2012} \\
8 &0.1993 &0.077 &  0.0051 &  \cite{Moresco/2012} \\
9 &0.2 & 0.077 & 0.030 &  \cite{Zhang/2014}\\
10 &0.24 & 0.075 &  0.0026 &  \cite{Gazta/2009}\\
11 &0.27 & 0.081 &  0.014 &  \cite{Stern/2010} \\
12 &0.28 & 0.079 &  0.035 &  \cite{Zhang/2014}\\
13 &0.35 & 0.091 &  0.0085 &  \cite{Chuang/2013}\\
14 &0.352 &0.085 &  0.0143 &  \cite{Moresco/2012} \\
15 &0.38 &0.085 &  0.0019 & \cite{Alam/2016} \\
16 &0.3802 &0.083 & 0.0137 &  \cite{Moresco/2016} \\
17 &0.4& 0.097 &  0.0173 &  \cite{Simon/2005} \\
18 &0.4004 &0.079 &  0.0104 &  \cite{Moresco/2016} \\
19 &0.4247 &0.089 &  0.0114 &  \cite{Moresco/2016} \\
20 &0.43 & 0.088 &  0.0038 &  \cite{Gazta/2009}\\
21 &0.44 & 0.084 &  0.008 &  \cite{Blake/2012}\\
22 &0.4449 & 0.095 & 0.013& \cite{Moresco/2016}\\
23 &0.47 & 0.091 &  0.051 & \cite{Ratsimbazafy/2017}\\
24 &0.4783 &0.083 &  0.009 &  \cite{Moresco/2016} \\
25 &0.48 & 0.099 &  0.061 &  \cite{Stern/2010} \\
26 &0.51 &0.092 &  0.0019 &  \cite{Alam/2016} \\
27 &0.57 &0.106 &  0.0035 &  \cite{Anderson/2014} \\
28 &0.593 &0.106 & 0.0132 &  \cite{Moresco/2012} \\
29 &0.6 & 0.089 & 0.0062 &  \cite{Blake/2012}\\
30 &0.61 &0.099 &  0.0021 &  \cite{Alam/2016} \\
31 &0.68 &0.094 &  0.0082 &  \cite{Moresco/2012} \\
32 &0.73 & 0.099 &  0.0072 &  \cite{Blake/2012}\\
33 &0.781 & 0.107 &  0.012 & \cite{Moresco/2012} \\
34 &0.875 & 0.128 &  0.0173 &  \cite{Moresco/2012} \\
35 &0.88 & 0.092 & 0.041 & \cite{Stern/2010} \\
36 &0.9 & 0.120 &  0.0234 &  \cite{Stern/2010} \\
37 &1.037 &0.157 &  0.020 &  \cite{Moresco/2012} \\
38 &1.3 & 0.172 &  0.0173 &  \cite{Stern/2010} \\
39 &1.363 &0.164 &  0.0343 &  \cite{Moresco/2015} \\
40 &1.43 & 0.181 &  0.0183 &  \cite{Stern/2010} \\
41 &1.53 & 0.143 &  0.0143 &  \cite{Stern/2010} \\
42 &1.75 & 0.207 &  0.041 &  \cite{Stern/2010} \\
43 &1.965 &0.191 &  0.0514 & \cite{Moresco/2015} \\
44 &2.3 & 0.229 &  0.0082 &  \cite{Busca/2013} \\
45 &2.34 & 0.227 &  0.0072 &  \cite{Delubac/2015} \\
46 &2.36 &0.231 &  0.0082 &  \cite{Ribera/2014} \\
\hline
\hline
\end{tabular}}
\label{tab:1}
\end{table}
Note that in the above References, $H(z)$ and error $\sigma_{i}$ are in the unit of $km~s^{-1}~Mpc^{-1}$. In this paper, we have converted these quantities in the unit of $Gyr^{-1}$.\\

\begin{figure}[h!]
\includegraphics[width=10.5cm,height=11cm,angle=0]{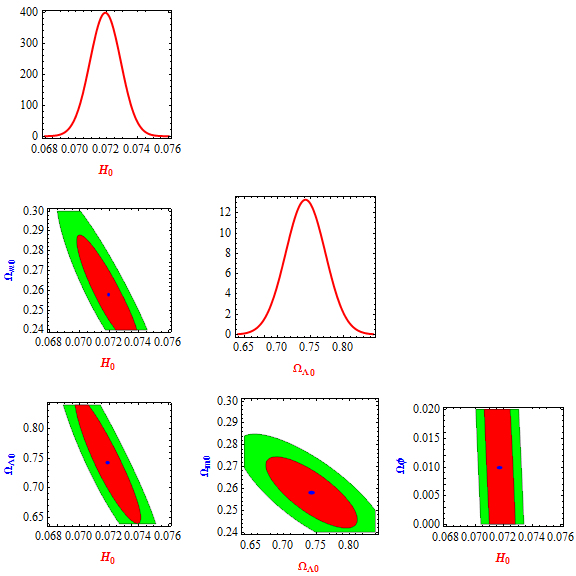}
\caption{One-dimensional marginalized distribution and two-dimensional contours with $68\%$ CL and $95\%$ CL for  parameter space {\bf$\Theta_{\sigma BD}$} using {\bf H(z)} data.}
\label{fig1}
\end{figure}
\begin{figure}[h!]
\includegraphics[width=10.5cm,height=11cm,angle=0]{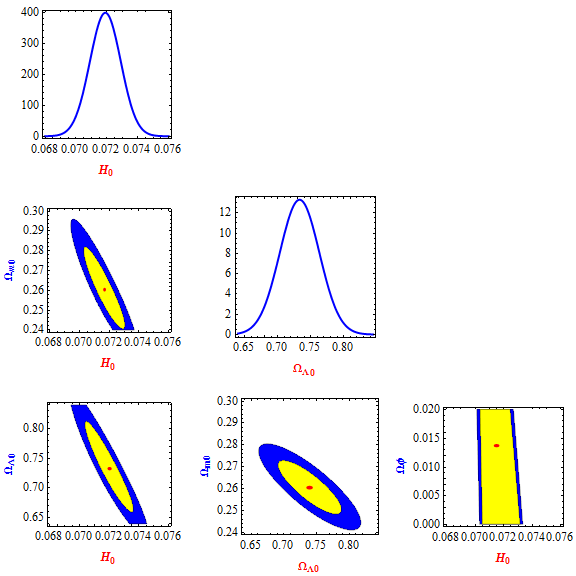}
\caption{One-dimensional marginalized distribution and two-dimensional contours with $68\%$ CL and $95\%$ CL for  parameter space {\bf$\Theta_{\sigma BD}$} using {\bf H(z)+ BAO} data.}
\label{fig2}
\end{figure}
\begin{figure}[h!]
\includegraphics[width=10.5cm,height=8.5cm,angle=0]{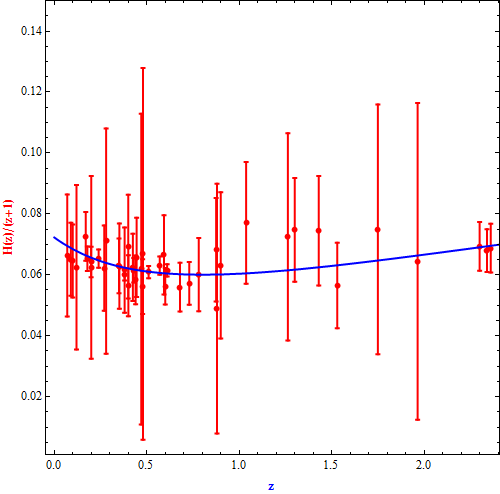}
\caption{The plot of Hubble rate versus the red-shift z. The points
with error bars indicate the experimental data summarized in Table \ref{tab:1}. H(z) is in unit of $Gyr^{-1}$.}
\label{fig3}
\end{figure}  
For all analysis, we have defined a $\chi^{2}$ for parameters with the likelihood given by $\zeta \propto e^{-\frac{\chi^2}{2}}$. Therefore, the $\chi^{2}$ function for $H(z)$ data is written as
\begin{equation}
\label{chi-1}
\chi^{2}_{ H}=\sum_{i=1}^{46}\left[\frac{H(z_{i},s)-H_{obs}(z_{i})}{\sigma_{i}}\right]^{2}
\end{equation}
where $s$ and $\sigma_{i}$ denote the parameter vector and standard error in in experimental values of Hubble's function $H$ respectively. \\
Similarly, the joint $\chi^{2}$ is read as
\begin{equation}
\label{chi-2}
\chi^{2}_{joint} = \chi^{2}_{H} + \chi^{2}_{BAO}
\end{equation}
Figures 1 and 2 exhibit the one-dimensional marginalized distribution and two-dimensional contours with $68\%$ CL and $95\%$ CL for parameter space {\bf$\Theta_{\sigma BD}$} using {\bf H(z)} and combined {\bf H(z)+ BAO} data respectively. The numerical result of statistical analysis is listed in table \ref{tab:2}.\\
\begin{table}[ht]
\caption{Summary of statistical analysis}
\centering
\setlength{\tabcolsep}{18pt}
\scalebox{1.22}{
\begin{tabular} {ccc}
\hline
\hline
Model parameters & $H(z)$  & $H(z)+BAO $\\[0.5ex] 

\hline{\smallskip}
$H_{0}$ & $0.0719 (Gyr^{-1})$ & $0.0717 (Gyr^{-1})$\\
$\Omega_{m0}$ & 0.258  &  0.261\\
$\Omega_{\Lambda 0}$ & 0.742 & 0.733\\
$\Omega_{\phi}$  & 0.0098  & 0.014\\
$\chi^{2}_{min}$ & 24.343 & 38.779\\
$\chi^{2}_{\nu}$  & 0.579 & 0.745 \\
\hline
\hline
\end{tabular}}
\label{tab:2}
\end{table}
We have summarized the numerical result of statistical analysis in table \ref{tab:2}. From table \ref{tab:2}, it has been observed that the estimated constraints on $H_{0}$ as $0.0719~Gyr^{-1}$ \\
$\sim$~$70.4$~$km~s^{-1}~Mpc^{-1}$ and $0.0717~Gyr^{-1}\sim~70.2~km~s^{-1}~Mpc^{-1}$ are closer to other investigations \cite{Chen/2011,Aubourg215,Chen/2017,Hinshaw/2013}. The best fit curve of Hubble rate versus redshift of derived model is shown in Fig. \ref{fig3}. In this paper, our aim is also to constrain the density of scalar field $\Omega_{\phi}$ and the estimated constraints on $\Omega_{\phi}$ with {\bf H(z)} and {\bf H(z)+ BAO} data are as $\Omega_{\phi}=0.0098$ and $\Omega_{\phi}=0.014$ respectively. In Amirhashchi and Yadav \cite{Amirhashchi/2019}, we also find constraint on scalar field density as $\Omega_{\phi} = 0.010$ by using different observational data sets. In table \ref{tab:2}, $\chi^{2}_{\nu}$ is read as $\chi^{2}_{\nu}=\chi^{2}_{min}/dof$ where dof is abbreviation of degree of freedom and it is defined as the difference between all observational data points and the number of free parameters. It should be noted that for $\chi^{2}_{\nu}\leq 1$, the fitting of model with observed data is considered as the best fitting model.\\     
\section{Properties of the model}
\subsection{The deceleration parameter}
\begin{figure}[h!]
\includegraphics[width=10cm,height=8cm,angle=0]{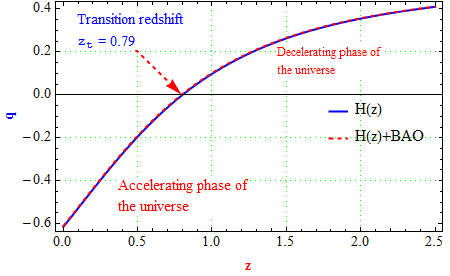}
\caption{The plot of deceleration parameter versus the red-shift z. The transition redshift is $z_{t} = 0.79$.}
\label{fig4}
\end{figure}  
The deceleration parameter in terms of redshift is read as
\begin{equation}
\label{q}
q(z) = -1+\frac{(1+z)H^{\prime}_{\sigma BD}}{H_{\sigma BD}}
\end{equation}
Here, $H^{\prime}_{\sigma BD}$ denotes first derivative of $H_{\sigma BD}$ with respect to $z$.\\
 
Using equation (\ref{H-BD}), equation (\ref{q}) is recast as
\begin{equation}
\label{q-1}
q(z)=-1+\frac{(z+1)\left(2.5 \Omega _{\text{m0}} \left(1. \sqrt{1-\Omega _{\phi }}+0.2\right) (z+1)^{2.5 \sqrt{1-\Omega _{\phi }}-0.5}+2 (z+1) \Omega _{\text{$\alpha $0}}+q_{1}\right)}{2\left(\Omega _{\text{$\Lambda $0}}+\Omega _{\text{m0}} (z+1)^{2.5 \sqrt{1-\Omega _{\phi}}+0.5}+(z+1)^2 \Omega _{\text{$\alpha $0}}+\Omega _{\text{$\sigma $0}} (z+1)^{5 \sqrt{1-\Omega _{\phi }}+1}\right)}
\end{equation}
where $q_{1} = \Omega _{\text{$\sigma $0}} \left(5 \sqrt{1-\Omega _{\phi }}+1\right) (z+1)^{5 \sqrt{1-\Omega _{\phi }}}$.\\
The present value of deceleration parameter is obtained as
\begin{equation}
\label{q-2}
q_{0} = -1+\frac{2 \Omega _{\text{$\alpha $0}}+\Omega _{\text{m0}} \left(2.5 \left(\sqrt{1-\Omega _{\phi }}-1\right)+3\right)+\Omega _{\text{$\sigma $0}} \left(5 \left(\sqrt{1-\Omega _{\phi }}-1\right)+6\right)}{2 \left(\Omega _{\text{$\alpha $0}}+\Omega _{\text{$\Lambda $0}}+\Omega _{\text{m0}}+\Omega _{\text{$\sigma $0}}\right)}
\end{equation}

The dynamics of deceleration parameter with the age of universe id depicted in Fig. \ref{fig4}. The derived model represents a transitioning universe with a transition redshift of about $z_{t} = 0.79$. We observe that the current universe is in accelerating phase while it was in decelerating phase of expansion in past. The present value of deceleration parameter $q_{0}$ is about $-0.61$. This value of $q_{0}$ is in excellent agreement with recent observations. 
\subsection{The age of universe}
The age of universe is obtained as
\begin{equation}
\label{age-1}
dt = -\frac{dz}{(1+z)H_{\sigma BD}}\Rightarrow 
\int_{t}^{t_{0}} dt  = - \int_{z}^{0}\frac{1}{(1+z)H_{\sigma BD}}dz
\end{equation}
Equations (\ref{H-BD})and equation (\ref{age-1}) lead to
\begin{equation}
\label{age-2}
 t_0 - t ={\int_{0}^{z}}\frac{(1-\Omega_{\phi})^\frac{1}{2}dz}{H_{0}(1+z)\left[\Omega_{m0}(1+z)^{2.5 \left(\sqrt{1-\Omega \phi }-1\right)+3}+\Omega_{\sigma 0}(1+z)^{5 \left(\sqrt{1-\Omega \phi }-1\right)+6}+ \Omega_{\alpha}(1+z)^{2}+\Omega_{\Lambda 0}\right]^\frac{1}{2}}
\end{equation}
Here, $t_{0}$ is the present age of the universe. Hence
\begin{equation}
\label{age-2}
 t_0 =\lim_{x\rightarrow\infty}{\int_{0}^{z}}\frac{(1-\Omega_{\phi})^\frac{1}{2}dz}{H_{0}(1+z)\left[\Omega_{m0}(1+z)^{2.5 \left(\sqrt{1-\Omega \phi }-1\right)+3}+\Omega_{\sigma 0}(1+z)^{5 \left(\sqrt{1-\Omega \phi }-1\right)+6}+ \Omega_{\alpha}(1+z)^{2}+\Omega_{\Lambda 0}\right]^\frac{1}{2}}
\end{equation}
Integrating equation (\ref{age-2}), we get
\begin{equation}
\label{age-3}
H_{0}t_{0} = 0.977621
\end{equation}
\begin{figure}[h!]
\includegraphics[width=10cm,height=8cm,angle=0]{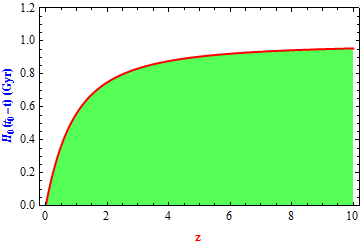}
\caption{The plot of $H_{0}(t_{0}-t)$ versus the red-shift z for $\Omega_{m0} = 0.261$, $\Omega_{\Lambda} = 0.733$ and $\Omega_{\phi} = 0.014$.}
\label{fig5}
\end{figure}  
From equation (\ref{age-3}), the present age of the universe is read as $t_{0} = 0.977621 H_{0}^{-1}$ $\sim$ 13.65 Gyrs. The plot of $H_{0}(t_{0}-t)$ versus redshift $z$ is graphed in Fig. \ref{fig5}. From Fig. \ref{fig5}, we observe that at present time $i. e.$ for $z = 0$, $H_{0}(t_{0}-t)$ is null which turn into imply $t = t_{0}$.  
\subsection{The particle horizon}
The particle horizon is the furthest distance from which one can retrieve information from the past, and hence defines the observable universe \cite{Bentabol/2013}. Thus the particle horizon is represented by proper distance measured by light signal coming from $t = 0$ to $t = t_{0}$.\\

Here, we assume light signal emits from a source along x-axis. The proper distance of
the source will be $a_{0}x$ and we are receiving that signal at present time $t_{0}$. Thus, the proper distance of the source from us is calculated as $a_{0}\int_{t_{p}}^{t_{0}}\frac{dt}{a(t)}$ where $t_{p}$ is the time in past at which the light signal was transmitted from source.\\
\begin{figure}[h!]
\includegraphics[width=10cm,height=8cm,angle=0]{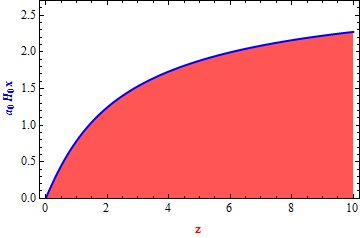}
\caption{The plot of proper distance $a_{0}H_{0}x$ versus the redshift $z$ for $\Omega_{m0} = 0.261$, $\Omega_{\Lambda} = 0.733$ and $\Omega_{\phi} = 0.014$.}
\label{fig6}
\end{figure}  

Therefore, the particle horizon is computed as
\begin{equation}
\label{PH}
R_{p} = lim_{t_{p}\rightarrow 0}\;\;a_{0}\int_{t_{p}}^{t_{0}}\frac{dt}{a(t)} = lim_{z\rightarrow \infty}\int_{0}^{z}\frac{dz}{H_{\sigma BD}}
\end{equation}
Using equation (\ref{H-BD}), equation (\ref{PH}) becomes
\begin{equation}
\label{PH-1}
R_{p}= lim_{z\rightarrow \infty}\int_{0}^{z}\frac{(1-\Omega_{\phi})^\frac{1}{2}dz}{H_{0}\left[\Omega_{m0}(1+z)^{2.5 \left(\sqrt{1-\Omega \phi }-1\right)+3}+\Omega_{\sigma 0}(1+z)^{5 \left(\sqrt{1-\Omega \phi }-1\right)+6}+ \Omega_{\alpha}(1+z)^{2}+\Omega_{\Lambda 0}\right]^\frac{1}{2}}
\end{equation}
Integrating equation (\ref{PH-1}) for $\Omega_{m0} = 0.261$, $\Omega_{\Lambda} = 0.733$ and $\Omega_{\phi} = 0.014$, we obtain
\begin{equation}
\label{Ph-2}
R_{p} = \frac{2.668}{H_{0}}
\end{equation}
Fig. \ref{fig6} shows variation of proper distance versus redshift. From Fig. \ref{fig6}, we observe that at present $i. e.$ for $z =0$, $a_{0}H_{0}x$ is null which turn into imply that $x\rightarrow \infty$. Thus we are at infinite distance from the first event occurred in past.\\
\subsection{The jerk parameter}
\begin{figure}[h!]
\includegraphics[width=10cm,height=8cm,angle=0]{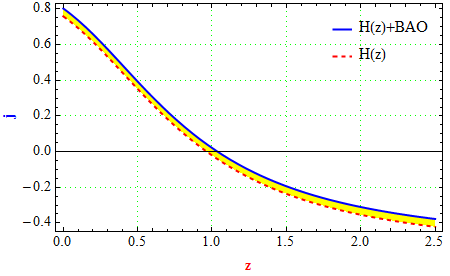}
\caption{The variation of jerk parameter versus redshift.}
\label{fig7}
\end{figure} 
The jerk parameter (j) \cite{Mukherjee/2019}, in terms of red-shift is given by 
\begin{equation}
\label{jerk-1}
j = 1-(1+z)\frac{H^{'}_{\sigma BD}}{H_{\sigma BD}}+\frac{1}{2}(1+z)^{2}\frac{[H^{''}_{\sigma BD}]^{2}}{[H_{\sigma BD}]^{2}}
\end{equation}
Equations (\ref{H-BD}) and (\ref{jerk-1}) lead to
\begin{equation}
\label{jerk-2}
j=1-(1+z)\xi_{1}+(1+z)^{2}\xi_{2}
\end{equation}
where\\

$\xi_{1}=\frac{2.5 \Omega _{m0} \left( \sqrt{1-\Omega _{\phi }}+0.2\right) (z+1)^{2.5 \sqrt{1-\Omega _{\phi 0}}-0.5}+2 (z+1) \Omega _{\alpha 0}+\Omega _{\sigma 0} \left(5 \sqrt{1-\Omega _{\phi }}+1\right) (z+1)^{5 \sqrt{1-\Omega _{\phi }}}}{2 \sqrt{1-\Omega _{\phi }} \sqrt{\Omega _{\Lambda 0}+\Omega _{m 0} (z+1)^{2.5 \sqrt{1-\Omega _{\phi }}+0.5}+(z+1)^2 \Omega _{\alpha 0}+\Omega _{\sigma 0} (z+1)^{5 \sqrt{1-\Omega _{\phi }}+1}}}$\\

$\xi_{2}=\frac{(w_{2}w_{3}-w_{1})^{2}}{w_{4}}$\\

$w_{1}=\frac{6.25 \left(\Omega _{m 0} \left(1. \sqrt{1-\Omega _{\phi }}+0.2\right) (z+1)^{2.5 \sqrt{1-\Omega _{\phi }}}+(z+1)^{0.5} \left((0.8 z+0.8) \Omega _{\alpha 0}+\Omega _{\sigma 0} \left(2. \sqrt{1-\Omega _{\phi }}+0.4\right) (z+1)^{5 \sqrt{1-\Omega _{\phi }}}\right)\right){}^2}{(z+1)}$\\

$w_{2}=\frac{4 \left(\Omega _{\Lambda 0}+\Omega _{m 0} (z+1)^{2.5 \sqrt{1-\Omega _{\phi }}+0.5}+(z+1)^2 \Omega _{\alpha 0}+\Omega _{\sigma 0} (z+1)^{5 \sqrt{1-\Omega _{\phi }}+1}\right)}{(z+1)^{1.5} (z+1)}$\\

$w_{3}=\Omega _{m 0} w_{5}(z+1)^{2.5 \sqrt{1-\Omega _{\phi }}}+(z+1)^{1.5} \left((z+1) \Omega _{\alpha 0}+\Omega _{\sigma 0} \left(-12.5 \Omega _{\phi }+2.5 \sqrt{1-\Omega _{\phi }}+12.5\right) (z+1)^{5 \sqrt{1-\Omega _{\phi }}}\right)$\\

$w_{4}=32 w_{6}\left(\Omega _{\Lambda 0}+\Omega _{m 0} (z+1)^{2.5 \sqrt{1-\Omega _{\phi }}+0.5}+(z+1)^2 \Omega _{\alpha 0}+\Omega _{\sigma 0} (z+1)^{5 \sqrt{1-\Omega _{\phi }}+1}\right)^{3}$\\

$w_{5}=(-3.125 z-3.125)\Omega _{\phi }+3 z+3$\\

$w_{6}=\left(\Omega _{\Lambda 0}+\Omega _{m 0} (z+1)^{2.5 \left(\sqrt{1-\Omega _{\phi }}-1\right)+3}+(z+1)^2 \Omega _{\alpha 0}+\Omega _{\sigma 0} (z+1)^{5 \left(\sqrt{1-\Omega _{\phi }}-1\right)+6}\right)$\\

In 2004, Blandford et al. \cite{Blandford/2004} have described the features of the jerk parameterization which gives an alternative approach to describe cosmological models close to $\Lambda$CDM model. A powerful feature
of of the jerk parameter is that for the ΛCDM model $j = 1$. In Refs. \cite{Sahni/2003,Alam/2003}, the authors have investigated the important features of $j$ for discriminating different dark energy models. The value $j \neq 1$ would favor a non-$\Lambda$CDM model. In the considered model, the explicit behavior of $j$ is shown in Fig. \ref{fig7}. We observe that the jerk parameter of considered model does not have $j = 1$. \\
\subsection{Shear scalar \& relative anisotropy}
The shear scalar is read as
\begin{equation}
\label{ss}
\sigma^{2} = \frac{1}{2}\sigma_{ij}\sigma^{ij}
\end{equation}
where $\sigma_{ij} = u_{i;j}-\theta(g_{ij}-u_{i}u_{j})$\\

In derived model, the shear scalar is given by
\begin{equation}
\label{ss-1}
\sigma^{2} = \frac{\dot{D}^{2}}{D^{2}} = k^{2}(1+z)^{6+5\sqrt{1-\Omega_{\phi}}-1}
\end{equation}
Thus the relative anisotropy is obtained as
\begin{equation}
\label{ra}
A_{m} = \frac{\sigma^{2}}{\rho_{m}}
\end{equation}
\begin{figure}\label{fig2}
\begin{center}
\includegraphics[width=10cm,height=8cm,angle=0]{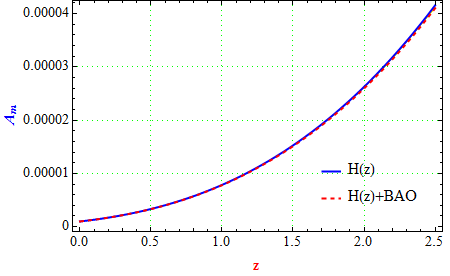}
\end{center}
\caption{The plot of relative anisotropy $A_{m}$ versus $z$.}
\end{figure}
From equation (\ref{ra}), it is clear that relative anisotropy depends on red-shift $z$. For high value of red-shift, the relative anisotropy is large and it decreases as with low value of $z$ and finally becomes null at $z\rightarrow 0$. This behavior of relative anisotropy $A_{m}$ is depicted in Fig. 8.\\    

\section{Concluding remarks}
In this paper, we have investigated a transitioning model of an-isotropic universe in Brains-Dicke theory of gravitation. We describe that the current phase of accelerated expansion of the universe is due to contribution coming from $\Lambda$ screened scalar field and the transition redshift is $z_{t} = 0.79$. For redshift $z > z_{t}$, the universe was in decelerating phase of expansion. Some important features of derived model are as follows:\\
\begin{itemize}
\item[i)] The derived model obeys Mach's principle.\\
\item[ii)] We find constraints on $H_{0}$, $\Omega_{m 0}$ and $\Omega_{\Lambda 0}$ by bounding the model under consideration with recent OHD and BAO data. The best fit values of $H_{0}$ are closer to other investigations \cite{Chen/2011,Aubourg215,Chen/2017,Hinshaw/2013}. Thus, we conclude the present OHD and BAO data provides well constrained values of $H_{0}$ and our model have good consistency with recent observations.\\
\item[iii)] We have estimated the present age of universe as $t_{0} = 13.65$ Gyrs. This age of universe is nicely matches with those obtained by Plank collaboration.\\
\item[iv)] The dynamics of deceleration parameter is showing a signature flipping from early decelerating phase to current accelerating phase at $z_{t} = 0.79$. The present value of deceleration parameter is computed as $q_{0} = -0.61$.\\
\item[v)] In the derived model, particle horizon exists and its value is different from $\Lambda$CDM model of universe.\\
\item[vi)] In the derived model, $j \neq 1$. Therefore, the derived solution describes the model of universe other than $\Lambda$CDM and the deviation from j = 1 investigates the dynamics of different kinds of dark energy
models other than $\Lambda$CDM. Some important applications of non $\Lambda$CDM model of the universe are given in Refs. \cite{Akarsu/2012,Singh/2019}. \\  
\end{itemize}
        

\end{document}